# Performance Guarantees for Distributed Reachability Queries

Wenfei Fan[1,2]    Xin Wang[1]    Yinghui Wu[1,3]
[1]University of Edinburgh    [2]Harbin Institute of Technology    [3]UC Santa Barbara
{wenfei@inf, x.wang-36@sms, y.wu-18@sms}.ed.ac.uk

## ABSTRACT

In the real world a graph is often fragmented and distributed across different sites. This highlights the need for evaluating queries on distributed graphs. This paper proposes distributed evaluation algorithms for three classes of queries: *reachability* for determining whether one node can reach another, *bounded reachability* for deciding whether there exists a path of a bounded length between a pair of nodes, and *regular reachability* for checking whether there exists a path connecting two nodes such that the node labels on the path form a string in a given regular expression. We develop these algorithms based on *partial evaluation*, to explore parallel computation. When evaluating a query $Q$ on a distributed graph $G$, we show that these algorithms possess the following *performance guarantees*, no matter how $G$ is fragmented and distributed: (1) each site is visited *only once*; (2) the total network traffic is determined by the size of $Q$ and the fragmentation of $G$, *independent of* the size of $G$; and (3) the response time is decided by the largest fragment of $G$ *rather than* the entire $G$. In addition, we show that these algorithms can be readily implemented in the MapReduce framework. Using synthetic and real-life data, we experimentally verify that these algorithms are scalable on large graphs, regardless of how the graphs are distributed.

## 1. INTRODUCTION

Large real-life graphs are often fragmented and stored distributively in different sites, *e.g.,* social networks [27], Web services networks [23] and RDF graphs [16, 26]. For instance, a graph representing a social network may be distributed across different servers and data centers for performance, management or data privacy reasons [12, 23, 25, 27] (*e.g.,* social graphs of Twitter and Facebook are geo-distributed to different data centers [12, 25]). Moreover, various data of people (*e.g.,* friends, products, companies) are typically found in different social networks [27], and have to be taken together when one needs to find the complete information about a person. With this comes the need for effective tech-niques to query distributed graphs, for *e.g.,* computing recommendations [17] and social network aggregations [27].

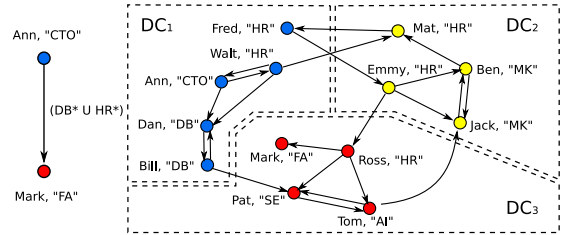

**Figure 1: Querying a distributed social network**

There have been a number of algorithms and distributed graph database systems for evaluating queries on distributed graphs (*e.g.,* [3, 6, 11, 29, 30]). However, few of these algorithms and systems provide *performance guarantees*, on the number of visits to each site, network traffic (data shipment) or computational cost (response time). The need for developing efficient distributed evaluation algorithms with performance guarantees is particularly evident for reachability queries, which are most commonly used in practice.

This paper advocates to evaluate queries on distributed graphs based on *partial evaluation*. Partial evaluation (*a.k.a.* program specialization) has been proved useful in a variety of areas including compiler generation, code optimization and dataflow evaluation (see [18] for a survey). Intuitively, given a function $f(s, d)$ and part of its input $s$, partial evaluation is to specialize $f(s, d)$ with respect to the known input $s$. That is, it conducts the part of $f$'s computation that depends only on $s$, and generates *a partial answer*, *i.e.,* a residual function $f'$ that depends on the as yet unavailable input $d$. This idea can be naturally applied to distributed query evaluation. Indeed, consider a query posed on a graph $G$ that is partitioned into fragments $(F_1, \ldots, F_n)$, where $F_i$ is stored in site $S_i$. To compute $Q(G)$, each site $S_i$ can find the partial answer to $Q$ in fragment $F_i$ *in parallel*, by taking $F_i$ as the known input $s$ while treating the fragments in the other sites as yet unavailable input $d$. These partial answers are collected and combined by a coordinator site, to derive the answer to query $Q$ in the entire $G$.

**Example 1:** Figure 1 depicts a fraction $G$ of a recommendation network, where each node denotes a person with name and job titles (*e.g.,* database researcher (DB), human resource (HR)), and each directed edge indicates a recommendation. The graph $G$ is *geo-distributed* to three data centers $DC_1$, $DC_2$ and $DC_3$, each storing a *fragment* of $G$.

Consider a query $Q$ given in Fig. 1, posed at $DC_1$. It is to find whether there exists a chain of recommendations from a CTO Ann to her finance analyst (FA) Mark, through either a list of DB people or a list of HR people. Observe that such a path exists: (Ann, CTO) → (Walt, HR) → (Mat, HR) → (Fred, HR) → (Emmy, HR) → (Ross, HR) → (Mark, FA). However, it is nontrivial to verify this in the distributed setting. A naive method is to first ship data from $DC_1$, $DC_2$ and $DC_3$ to a single site, and then evaluate the query using an algorithm developed for centralized data (*i.e.,* graphs stored in a single site). This is infeasible because its data shipment may be prohibitively expensive and worse still, may not even be allowed for data privacy. Another way is to use a distributed graph traversal algorithm, by sending messages between different sites. This, however, requires messages to be sent along $DC_1 \to DC_2 \to DC_1 \to DC_2 \to DC_3 \to DC_1$, incurring unbounded number of visits to each site, excessive communication cost, and unnecessary delay in response.

We can do better by using partial evaluation. We send the query $Q$ to $DC_1$, $DC_2$ and $DC_3$, as is. We compute the partial answers to (sub-queries of) $Q$ at each site, in parallel, by taking the fragment residing in the site as known input and introducing Boolean variables to indicate unknown input (*i.e.,* fragments in the other sites). The partial answers are vectors of Boolean formulas, one associated with each node that has an edge from a fragment stored at another site. These Boolean formulas indicate (1) at $DC_1$, from Ann there exist an HR path to Walt and a DB path to Bill, and from Fred there is an HR path to Emmy; (2) at $DC_2$, there exist an HR path from Emmy to Ross, an HR path from Mat to Fred; and (3) at $DC_3$, there exists an HR path from Ross to Mark. These partial answers are collected by a coordinator site ($DC_1$), which solves a system of equations formed by these Boolean formulas that are *recursively defined*, to find the truth values of those Boolean variables. It yields answer true to $Q$, *i.e.,* there exists an HR path from Ann to Mark.

We will show that this method guarantees the following: (1) each site is visited *only once*; (2) besides the query $Q$, only *2* messages are sent, all to the coordinator, and each message is *independent of* the size of $G$, and (3) partial evaluation is conducted *in parallel* at each site, *without waiting for* the outcome or messages from any other site. □

While there has been work on query answering via partial evaluation [2, 3, 6, 11], the previous work has focused on either trees [2, 3, 6] or non-recursive queries expressed in first-order logic (FO) [11]. We are not aware of any previous algorithms based on partial evaluation for answering reachability queries, which are *beyond* FO, on *possibly cyclic graphs* that are *arbitrarily* fragmented and distributed.

**Contributions.** We provide distributed evaluation algorithms for three classes of reachability queries commonly used in practice, via partial evaluation. We show that these algorithms posses several salient *performance guarantees*.

(1) Our first algorithm is developed for *reachability queries* (Section 3), to decide whether two given nodes are connected by a path [31]. We show that when evaluating such a query on a distributed graph $G$, the algorithm (a) visits each site *only once*, (b) is in $O(|V_f||F_m|)$ time, and (c) its total amount of data shipped is bounded by $O(|V_f|^2)$, where $|V_f|$ is the number of nodes that have edges across different sites, and $|F_m|$ is the size of the *largest fragment* in $G$.

(2) Our second algorithm is for evaluating *bounded reachability queries* (Section 4), for determining whether two given nodes are connected by a path of a bounded length [31]. We show that this algorithm has the same performance guarantees as its counterpart for reachability queries.

(3) Our third algorithm is to evaluate *regular reachability queries* (Section 5), to decide whether there exists a path between a pair $(u, v)$ of nodes such that the node labels on the path satisfy a regular expression $R$. When evaluating such a query on a distributed graph $G$, the algorithm (a) visits each site *only once*, (b) is in $O(|F_m||R|^2 + |R|^2|V_f|^2)$ time, and (c) has network traffic bounded by $(|R|^2|V_f|^2)$, where $|F_m|$ and $|V_f|$ are as above, and $|R|$ is the size of regular expression $R$, which is much smaller than $|V_f|$ and $|F_m|$.

(4) We also develop a MapReduce [7] algorithm for evaluating regular reachability queries (Section 6). This shows that partial evaluation can be readily implemented in the widely used MapReduce framework. The algorithm can be easily adapted to evaluate (bounded) reachability queries, which are special cases of regular reachability queries.

(5) We experimentally evaluate the efficiency and scalability of our algorithms(Section 7). We find that our algorithms scale well with both the size of graphs and the number of fragments. For instance, it takes 16 seconds to answer a regular reachability query on graphs with 1.5M (million) nodes and 2.1M edges, partitioned into 10 fragments. We also find that the communication cost of our algorithms is low. Indeed, the amount of data shipped by our algorithms is no more than 11% of the graphs in average. For reachability queries on real-life graphs, our algorithms take only 6% of running time of the algorithms based on message passing [21], and visit each site only once as opposed to 625 visits in average by its counterpart [21]. In addition, our MapReduce algorithm is efficient.

We contend that partial evaluation yields a promising approach to evaluating queries on distributed graphs. It guarantees that (1) the number of visits to each site is *minimum*; (2) the total network traffic is *independent of* the size of the entire graph; (3) the evaluation is conducted *in parallel*, and its cost depends on the largest fragment of a partitioned graph and the number of nodes with edges to different sites, *rather than the entire graph*; and (4) it imposes *no constraints* on how the graph is fragmented and distributed. Moreover, it can be readily implemented in the MapReduce model, as verified in our experimental study.

**Related Work.** We categorize related work as follows.

<u>Distributed databases</u>. A variety of distributed database systems have been developed. (1) Distributed relational databases (see [24]) can store graphs in distributed relational tables, but do not support efficient graph query evaluation [8, 9]. (2) Non-relational distributed data storage manage distributed data via various data structures, *e.g.,* sorted map [4], key/value pairs [8]. These systems are built forprimary-key only operations [8,9], or simple graph queries (*e.g.,* degree, neighborhood)[1], but do not efficiently support distributed reachability queries. (3) Distributed graph databases. Neo4j[1] is a graph database optimized for graph traversal. Trinity[2] and HyperGraphDB[3] are distributed systems based on hypergraphs. Unfortunately, they do not support efficient distributed (regular) reachability queries.

---

[1] *http://neo4j.org/*

[2] *http://research.microsoft.com/en-us/projects/trinity/*

[3] *http://www.kobrix.com/hgdb.jsp*

Closer to our work is Pregel [21], a distributed graph querying system based on message passing It partitions a graph into clusters, and selects a master machine to assign each part to a slave machine. A graph algorithm allows (a) the nodes in each slave machine to send messages to each other, and (b) the master machine to communicate with slave machines. Several algorithms (distance, etc.) supported by Pregel are addressed in [21]. Similar message-sending approaches are also developed in [13]. These algorithms differ from ours as follows. (a) In contrast to our algorithms, the message passing model in Pregel may serialize operations that can be conducted in parallel, and have no bound on the number of visits to each site, as shown by our experimental study (Section 7). (b) How to support regular reachability query is not studied in [21]. On the other hand, the techniques of Pregel can be combined with partial evaluation to support local processing of reachability queries at each site (see Section 3).

*Distributed graph query evaluation.* Several algorithms have been developed for evaluating queries on distributed graphs (see [19] for a survey). (1) Querying distributed trees [2,3,6]. Partial evaluation is used to evaluate XPath queries on distributed XML data modeled as trees [3, 6], as well as for evaluating regular path queries [2]. It is nontrivial, however, to extend these algorithms to deal with (possibly *cyclic*) graphs. Indeed, the network traffic of [3,6] is bounded by *the number of fragments* and the size of the query, in contrast to *the number of nodes* with edges to different fragments in our setting. Moreover, we study (regular) reachability queries, which are quite different from XPath. Finally, our algorithms only visit each site once, while in [2] each site may be visited multiple times. (2) Querying distributed semi-structured data [13, 28–30]. Techniques for evaluating regular path queries on distributed, edge-labeled, rooted graphs are studied in [30] and extended in [29], based on message passing. It is guaranteed that the total network traffic is bounded by $n^2$, where $n$ is the number of edges across different sites. A distributed BFS algorithm is given in [28], which takes nearly cubic time in graph size, and a table of exponential size to achieve a linear time complexity, and is impractical for large graphs. These differ from our algorithms as follows. (a) Our algorithms guarantee that each site is visited *only once*, as opposed to *twice* [30]. (b) As remarked earlier, message passing may unnecessarily serialize operations, while our algorithms explore parallelism via partial evaluation. While an analysis of computational cost is not given in [29, 30], We show experimentally that our algorithms outperform theirs (Section 7).

There has also been recent work on evaluating SPARQL queries on distributed RDF graphs [11], which is not applicaple to our setting due to (a) no performance guarantees or complexity bounds are provided in [11], and (b) the queries considered in [11] are expressible in FO, while we study (regular) reachability queries beyond FO.

## 2. DISTRIBUTED GRAPHS AND QUERIES

We start with distributed graphs (Section 2.1), reachability queries and a partial evaluation framework (Section 2.2).

### 2.1 Distributed Graphs

We start with basic notations of graphs. We consider node-labeled, directed graphs, simply referred to as graphs.

**Graphs**. A *graph* $G = (V, E, L)$ consists of (1) a finite set $V$ of nodes; (2) a set of edges $E \subseteq V \times V$, where $(v, w) \in E$ denotes a *directed* edge from node $v$ to $w$; and (3) a function $L$ defined on $V$ such that for each node $v$ in $V$, $L(v)$ is a label from a set $\Sigma$ of labels. Intuitively, $L()$ specifies node attributes, *e.g.,* names, keywords, social roles, ratings, companies [20]; the set $\Sigma$ specifies all such attributes.

We will use the following notations.
(1) A *path* $\rho$ from node $v$ to $w$ in $G$ is a sequence of nodes $(v = v_0, v_1, \ldots, v_n = w)$ such that for every $i \in [1, n]$, $(v_{i-1}, v_i) \in E$. The *length* of path $\rho$, denoted by $\mathsf{len}(\rho)$, is the number of edges in $\rho$. We define the *label* of $\rho$ to be the list of the labels of $v_1, \ldots, v_{n-1}$, excluding $v_0$ and $v_n$. Abusing notations of trees, we refer to $v_i$ as a *child* of $v_{i-1}$, and $v_j$ as a *descendant* of $v_i$ for $i, j \in [0, n]$ and $i < j$.

We say that a node $v$ can *reach* $w$ if and only if (iff) there is a path from $v$ to $w$. The *distance* from $v$ and $w$, denoted by $\mathsf{dist}(u, v)$, is the length of the shortest paths from $v$ to $w$.

(2) A *node induced subgraph* $G_s$ of $G$ is a graph $(V_s, E_s, L_s)$, where (a) $V_s \subseteq V$, (b) there is an edge $(u, v) \in E_s$ iff $u, v \in V_s$ and $(u, v) \in E$, and (c) for each $v \in V_s$, $L_s(v) = L(v)$.

**Distributed Graphs**. In practice a graph $G$ is often partitioned and stored in different sites [16, 27]. We define a *fragmentation* $\mathcal{F}$ of a graph $G = (V, E, L)$ as a pair $(F, G_f)$, where $F$ is a collection of subgraphs of $G$, and $G_f$ is called the *fragment graph* of $\mathcal{F}$, specifying edges across distinct sites. More specifically, $F$ and $G_f$ are defined as follows.

(1) $F = (F_1, \ldots, F_k)$, where each *fragment* $F_i$ is specified by $(V_i \cup F_i.O, E_i \cup cE_i, L_i)$ such that (a) $(V_1, \ldots, V_k)$ is a partition of $V$, (b) each $(V_i, E_i, L_i)$ is a subgraph of $G$ induced by $V_i$, (c) for each node $u \in V_i$, if there exists an edge $(u, v) \in E$, where $v$ is in another fragment, then there is a *virtual node* $v$ in $F_i.O$, and (d) $cE_i$ consists of all and only those edges $(u, v)$ such that $u \in V_i$ and $v$ is a virtual node, referred to as *cross edges*. We also use $F_i.I$ to denote the set of *in-nodes* of $F_i$, *i.e.,* those nodes $u \in V_i$ such that there exists a cross edge $(v, u)$ *incoming* from a node $v$ in another fragment $F_j$ to $u$, *i.e.,* $v$ is a virtual node in $F_j$.

Intuitively, $V_i \cup F_i.O$ of $F_i$ consists of (a) those nodes in $V_i$ and (b) for each node in $V_i$ that has an edge to another fragment, a virtual node indicating the connection. The edge set $E_i \cup cE_i$ consists of (a) the edges in $E_i$ and (b) *cross edges* in $cE_i$, *i.e.,* edges to other fragments. In a distributed social graph, for instance, cross edges are indicated by either IRIs (universal unique IDs) or semantic labels of the virtual nodes [21, 27]. We also identify $F_i.I$, a subset of nodes in $V_i$ to which there are incoming edges from another fragment.

We assume *w.l.o.g.* that each $F_i$ is stored at site $S_i$.

(2) The fragment graph $G_f$ is defined as $(V_f, E_f)$, where $V_f = \bigcup_{i \in [1,k]} (F_i.O \cup F_i.I)$ and $E_f = \bigcup_{i \in [1,k]} cE_i$. Here $F_i.O \cup F_i.I$ includes all the nodes in $F_i$ that have cross edges to or from fragment $F_i$. These nodes can be grouped together, denoted by a single "hyper-node", indicating $F_i$. The set $E_f$ collects all the cross edges from all fragments.

**Example 2:** Figure 1 depicts a fragmentation $\mathcal{F}$ of graph $G$, consisting of three fragments $F_1, F_2, F_3$ stored in sites $\mathsf{DC}_1$, $\mathsf{DC}_2$ and $\mathsf{DC}_3$, respectively. For fragment $F_1$, $F_1.O$ consists of virtual nodes Pat, Mat and Emmy, $F_1.I$ includes in-nodes Fred, and its $cE$ set consists of cross edges (Fred, Emmy), (Bill, Pat) and (Walt, Mat), *i.e.,* all the edges from $F_1$ outgoing to another fragment; similarly for $F_2$ and $F_3$. In particular, edges (Mat, Fred) and (Bill, Pat) are cross edges from fragments $F_2$ to $F_1$ and $F_1$ to $F_3$, respectively.

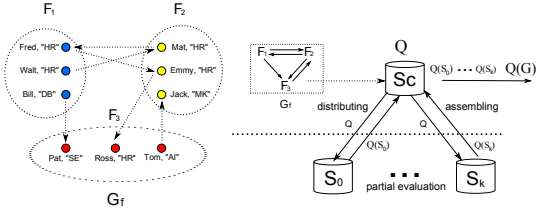

**Figure 2: Fragment graph and partial evaluation**

The fragment graph $G_f$ of $\mathcal{F}$ is shown in Fig. 2, which collects all in-nodes, virtual nodes and cross edges, but does not contain any nodes and edges internal to a fragment. □

We remark that *no constraints* are imposed on fragmentation, *i.e.*, the graphs can be *arbitrarily* fragmented. Observe that multiple fragments may reside in a single site, and our algorithms can be easily adapted to accommodate this.

## 2.2 Queries and Partial Evaluation

Given a fragmentation $\mathcal{F}$ of graph $G$ and a query $Q$, *distributed query evaluation* is to compute the answer to $Q$ in $G$, using data in $\mathcal{F}$. It aims to minimize (1) the number of visits to each site, (2) the network traffic (communication cost), *i.e.*, the total amount of data shipped from one site to another, and (3) the response time (computational cost).

We focus on three classes of graph queries in this work.
(1) A *reachability query* $\mathsf{q}_\mathsf{r}(s,t)$ is to determine whether node $s$ can reach another node $t$ in $G$.
(2) A *bounded reachability query* $\mathsf{q}_\mathsf{br}(s,t,l)$ is to decide whether $\mathsf{dist}(s,t) \leq l$ for a given integer (bound) $l$.
(3) A *regular reachability (path) query* $\mathsf{q}_\mathsf{rr}(s,t,R)$ is to determine whether there exists a path $\rho$ from $s$ to $t$ such that $\rho$ satisfies $R$. Here $R$ is a regular expression:

$$R ::= \epsilon \mid a \mid RR \mid R \cup R \mid R^*,$$

where $\epsilon$ is the empty string, $a$ is a label in $\Sigma$, $RR$ and $R \cup R$ and $R^*$ denote alternation, concatenation and the Kleene closure, respectively. We say that a path $\rho$ *satisfies* $R$ if the label of $\rho$ is a string in the regular language defined by $R$.

**Remark**. Observe the following. (1) One can define a "wildcard" $\_$, which matches any label, as $a_1 \cup \ldots \cup a_m$, for all $a_i$'s in $\Sigma$. Leveraging $\_$, reachability and bounded reachability queries can be expressed as regular reachability (path) queries. We study these queries separately because (a) they admit lower complexity than regular reachability queries, and (b) in practice, it often suffices to use these simple queries [31], without paying the price of higher complexity of regular path queries. (2) It is known that it is NP-complete to determine whether there exists a *simple* path $\rho$ from $s$ to $t$ such that $\rho$ satisfies a regular expression $R$ [22]. Here we do not require $\rho$ to be a simple path, *i.e.*, we allow multiple occurrences of the same node on $\rho$, and develop a low polynomial time algorithm for regular path queries.

Notations in this section are summarized in Table 1.

**Partial evaluation**. Given a query $Q$ and a fragmentation $\mathcal{F}$ of a graph $G$, we compute $Q(G)$, a Boolean value indicating the reachability of $Q$ in $G$. Assume that $Q$ is posed on a site $S_c$, referred to as a *coordinator site*, in which a mapping $h$ from the fragments in $\mathcal{F}$ to different sites is stored. As shown in Fig. 2, we use partial evaluation to compute $Q(G)$.

| symbols | notations |
|---|---|
| $\mathcal{F} = (F, G_f)$ | graph fragmentation in which $G_f$ is the fragment graph |
| $F_i.I$ | the set of in-nodes in a fragment $F_i$ |
| $F_i.O$ | the set of virtual nodes in a fragment $F_i$ |
| $\mathsf{q}_\mathsf{r}(s,t)$ | reachability query |
| $\mathsf{q}_\mathsf{br}(s,t,l)$ | bounded reachability query |
| $\mathsf{q}_\mathsf{rr}(s,t,R)$ | regular reachability query |

**Table 1: Notations: graphs and queries**

*(1) Distributing at site $S_c$.* Upon receiving $Q$, the coordinating site $S_c$ posts $Q$ to each fragment, as is, by using $h$.

*(2) Local evaluation at each site $S_i$.* Each site $S_i$ evaluates (sub-queries) of $Q$ *in parallel*, by treating the fragment $F_i$ stored in $S_i$ as the known input to $Q$; the other fragments $F_j$ are taken as the yet unavailable input, denoted by Boolean variables associated with virtual nodes in $F_i.O$. The partial answers are represented as vectors of Boolean formulas associated with nodes in $F_i.I$, and are sent back to $S_c$.

*(3) Assembling at $S_c$.* Site $S_c$ assembles these partial answers to get the final answer $Q(G)$, by using $G_f$.

Following this, the next three sections develop evaluation algorithms for (bounded, regular) reachability queries.

## 3. DISTRIBUTED REACHABILITY

We first develop distributed evaluation strategies for reachability queries. Given a reachability query $\mathsf{q}_\mathsf{r}(s,t)$ and a fragmentation $\mathcal{F} = (F, G_f)$ of a graph $G$, we decide whether $s$ reaches $t$ in $G$. The main result of this section is as follows.

**Theorem 1:** *Over a fragmentation $\mathcal{F} = (F, G_f)$ of a graph $G$, reachability queries can be evaluated (a) in $O(|V_f||F_m|)$ time, (b) by visiting each site only once, and (c) with the total network traffic bounded by $O(|V_f|^2)$, where $G_f = (V_f, E_f)$ and $F_m$ is the largest fragment in $F$.* □

As a proof of the theorem, we provide an algorithm to evaluate reachability queries $\mathsf{q}_\mathsf{r}(s,t)$ over a fragmentation $\mathcal{F}$ of a graph $G$. The algorithm, denoted as disReach, is given in Fig. 3. As shown in Fig. 2, the algorithm evaluates $\mathsf{q}_\mathsf{r}(s,t)$ based on partial evaluation, in three steps as follows.
(1) The coordinator site $S_c$ posts the same query $\mathsf{q}_\mathsf{r}(s,t)$ to each fragment in $F$ (line 1).
(2) Upon receiving $\mathsf{q}_\mathsf{r}(s,t)$, each site invokes procedure localEval to partially evaluate $\mathsf{q}_\mathsf{r}(s,t)$, *in parallel* (lines 3-4). This yields a *partial* answer $F_i.$rvset from each fragment, which is a set of Boolean equations (as will be discussed shortly) and is sent back to the coordinator site $S_c$.
(3) The coordinator site $S_c$ collects $F_i.$rvset from each site and assembles them into a system RVset of Boolean equations (line 3). It then invokes procedure evalDG to solve these equations and finds the final answer to $\mathsf{q}_\mathsf{r}(s,t)$ in $G$ (line 5). In contrast to partial query evaluation on trees [2, 3, 6], the Boolean equations of RVset are possibly *recursively defined* since graph $G$ may have a cyclic structure,

We next present procedures localEval and evalDG, for producing and assembling partial answers, respectively.

**Partial evaluation**. Procedure localEval evaluates $\mathsf{q}_\mathsf{r}(v,t)$ on each fragment $F_i$ in parallel. For each in-node $v$ in $F_i$, it decides whether $v$ reaches $t$. Later on procedure evalDG will assemble such answers and find the final answer to $\mathsf{q}_\mathsf{r}(s,t)$.

Let us consider how to compute $\mathsf{q}_\mathsf{r}(v,t)$. If $t \in F_i$ and $v$ can reach $t$, then $\mathsf{q}_\mathsf{r}(v,t)$ can be *locally evaluated* to be true. Otherwise, $\mathsf{q}_\mathsf{r}(v,t)$ is true iff there *exists* a virtual node $v'$ of $F_i$ such that *both* $\mathsf{q}_\mathsf{r}(v,v')$ and $\mathsf{q}_\mathsf{r}(v',t)$ are true. Indeed, in

Algorithm disReach   /* executed at the coordinator site */
Input: Fragmentation $(F, G_f)$, reachability query $q_r(s,t)$.
Output: The Boolean answer ans to $q_r(s,t)$ in $G$.

1. post query $q_r(s,t)$ to all the fragments in $F$;
2. RVset := $\emptyset$;
3. **for each** fragment $F_i$ in $F$ **do**
4.     RVset := RVset $\cup$ localEval($F_i, q_r(s,t)$);
5. ans := evalDG(RVset);
6. **return** ans;

**Procedure** localEval   /* locally at each site in parallel */
Input: A fragment $F_i$, a reachability query $q_r(s,t)$.
Output: (a set rvset of Boolean equations).

1. $F_i$.rvset := $\emptyset$; iset := $F_i.I$; oset := $F_i.O$;
2. **if** $s \in F_i$ **then** iset := iset $\cup \{s\}$;
3. **if** $t \in F_i$ **then** oset := oset $\cup \{t\}$;
4. **for each** node $v \in$ oset **do**
5.     **if** $v = t$ **then** $v$.rf := true;
6.     **else** $v$.rf := $X_v$;
7. **for each** node $v \in$ iset **do**
8.     **for each** node $v' \in$ oset **do**
9.         **if** $v' \in$ des$(v, F_i)$ **then** $v$.rf := $v$.rf $\vee v'$.rf;
10.    $F_i$.rvset := $F_i$.rvset $\cup \{X_v = v.\text{rf}\}$;
11. send $F_i$.rvset to the coordinator site $S_c$;

**Figure 3: Algorithm disReach**

the latter case $v$ can reach $t$ if there *exists* a virtual node $v'$ such that $v'$ can reach $t$. Observe that $q_r(v, v')$ can be *locally evaluated* in $F_i$, *but not* $q_r(v', t)$ since $v'$ and $t$ are in other fragments. Instead of waiting for the answer of $q_r(v', t)$, we introduce *Boolean variables*, one for each virtual node $v'$ in $F_i.O$, to denote the yet unknown answer to $q_r(v', t)$ in $G$. The answer to $q_r(v, t)$ is then a *Boolean formula* $v$.rf associated with $v$, which is the *disjunction* of *only* the variables of those virtual nodes $v'$ to which $v$ can reach in $F_i$.

More specifically, procedure localEval works as follows. It first initializes a set $F_i$.rvset of Boolean equations, and puts the in-nodes $F_i.I$ and virtual nodes $F_i.O$ of $F_i$ in sets iset and oset, respectively (line 1). If $s$ (resp. $t$) is in $F_i$, localEval includes $s$ (resp. $t$) in iset (resp. oset) as well (lines 2-3). A Boolean variable $X_v$ is associated with each node $v \in$ oset $\cup$ iset. For each virtual node $v \in$ oset, if $v$ is $t$ or $v$ can reach $t$ via a path in $F_i$, then $X_v$ is assigned true (lines 4-5). For each in-node $v \in$ iset, localEval *locally* checks whether $v$ can reach a virtual node $v' \in$ oset (lines 8-9). If so, localEval updates $v$.rf, the Boolean formula of $v$, to be $v$.rf $\vee v'$.rf (line 10). Observe that if $t$ is in des$(v, F_i)$, then $v$.rf is evaluated to be true. Here $v' \in$ des$(v, F_i)$ denotes that $v'$ is a descendant of $v$ in $F_i$; this can be checked using any available *centralized algorithm* for reachability queries [31], *locally* in $F_i$. After the formula of in-node $v$ is constructed, $F_i$.rvset is extended by including a *Boolean equation* $X_v = v$.rf. The set $F_i$.rvset is then sent to the coordinator site $S_c$ (line 11).

**Example 3:** Consider a query $q_r(\text{Ann}, \text{Mark})$ over $G$ in Fig 1. Algorithm disReach at the coordinator site $DC_1$ first sends the query to each site, where a set of Boolean equations are computed, as shown below.

| $F_i$ | $F_i.I$ | rf | rvset |
|---|---|---|---|
| $F_1$ | Ann | $x_{\text{Pat}} \vee x_{\text{Mat}}$ | $\{x_{\text{Ann}} = x_{\text{Pat}} \vee x_{\text{Mat}},\ x_{\text{Fred}} = x_{\text{Emmy}}\}$ |
|  | Fred | $x_{\text{Emmy}}$ |  |
| $F_2$ | Mat | $x_{\text{Fred}}$ | $\{x_{\text{Mat}} = x_{\text{Fred}},$ |
|  | Jack | $x_{\text{Fred}}$ | $x_{\text{Jack}} = x_{\text{Fred}},$ |
|  | Emmy | $x_{\text{Fred}} \vee x_{\text{Ross}}$ | $x_{\text{Emmy}} = x_{\text{Fred}} \vee x_{\text{Ross}}\}$ |
| $F_3$ | Ross | true | $\{x_{\text{Ross}} = \text{true},\ x_{\text{Pat}} = x_{\text{Jack}}\}$ |
|  | Pat | $x_{\text{Jack}}$ |  |

Observe that for each $i \in [1, 3]$, each equation in $F_i$.rvset is of the form $X_v = \bigvee X_{v'}$, where $v$ is an in-node, and $v'$ is

**Procedure** evalDG   /* executed at the coordinator site */
Input: A system RVset of Boolean equations.
Output: The Boolean answer ans to $q_r(s,t)$.

1. construct dependency graph $G_d = (V_d, E_d, L_d)$ from RVset;
2. **if** there is no $v_d \in V_d$ such that $L(v_d) = \{X_v = \text{true}\}$
    **then return** false;
3. **else** merge all such nodes into a node $v_{\text{true}}$;
4. **if** $v_{\text{true}} \in$ des$(v_s, G_d)$ **then return** true;
5. **else return** false;

**Figure 4: Procedure evalDG**

a virtual node that $v$ can reach in $F_i$. In particular, Ross.rf = true since the node Ross can reach Mark in $F_3$. □

**Assembling**. After the local evaluation, the equations collected in RVset at the coordinator site $S_c$ form a *Boolean equation system* (BES) [14]. It consists of equations of the form $X_v = v$.rf, where $v$ is an in-node in some fragment $F_i$, and Boolean variables in $v$.rf are associated with virtual nodes (out-nodes), which in turn are connected to in-nodes of some other fragments. In particular, RVset contains a Boolean equation $X_s = s$.rf, where the truth value of $X_s$ is the final answer to $q_r(s,t)$. Given RVset, procedure evalDG is to compute the truth value of $X_s$. Observe that equations in RVset may be defined *recursively*. For example, $x_{\text{Fred}}$ in Example 3 is defined indirectly in terms of itself.

Observe that RVset has $O(|V_f|)$ Boolean equations. It is known that BES RVset can be solved in $O(|V_f|^2)$ time [14]. We next present such an algorithm, based on a notion of dependency graphs. The *dependency graph* of RVset is defined as $G_d = (V_d, E_d, L_d)$, where $v_d \in V_d$ is a Boolean variable $X_v$ in RVset; $L_d(v_d) = \bigvee X_{v_i}$ if $X_v = \bigvee X_{v_i}$ is in RVset; and there is an edge $(v_d, v_d') \in E_d$ if and only if $X_v'$ is in $\bigvee X_{v_i}$ of $L_d(v_d)$. Note that the size $|G_d|$ of $G_d$ is in $O(|V_f|^2)$, where $G_f = (V_f, E_f)$ is the fragment graph of $\mathcal{F}$.

Based on this notion, we present procedure evalDG in Fig 4. It first constructs the dependency graph $G_d$ of RVset (line 1). It groups into a single node $v_{\text{true}}$ all those nodes (variables) that are known to be true (line 3). It returns false if no such node exists, since no in-nodes can reach $t$ in any of the fragment (line 2). Otherwise, it returns true if $v_s$ (indicating $X_s$ in $X_s = s$.rf) can reach $v_{\text{true}}$ (lines 4-5).

**Example 4:** Consider the Boolean equations of Example 3. Given these, evalDG first builds its dependency graph, shown in Fig 5(a). It then checks whether there is a path from $X_{\text{Ann}}$ to $X_{\text{true}}$ ($X_{\text{Mark}}$). It returns true as such a path exists. □

**Correctness**. One can easily verify the following: $s$ can reach $t$ in $G$ iff there exist a positive integer $l$ and a path $(s, x_1, \ldots, x_l, t)$ such that $x_i$.rf's are built in some fragment by localEval, and moreover, are evaluated to true by procedure evalDG. This can be shown by induction on $l$.

**Complexity**. Algorithm disReach guarantees the following.

*The number of visits.* Obviously each site is visited only once, when the coordinator site posts the input query.

*Total network traffic.* For each fragment $F_i$, $F_i$.rvset has $|F_i.I|$ equations, each of $|F_i.O|$ bits indicating the presence or absence of variables in the Boolean formula. Hence the set RVset consists of at most $|V_f|$ equations, each of at most $|V_f|$ bits. The total network traffic is thus bounded by $O(|V_f|^2)$, independent of $|G|$, since $|q_r(s,t)|$ is negligible.

*Computational cost.* Observe the following. (1) Procedure localEval is performed on each fragment $F_i$ *in parallel*, and it takes $O(|F_i||V_f|)$ time to compute $F_i$.rvset for each fragment (see the discussion below). Hence it takes at most $O(|V_f||F_m|)$ time to get $F_i$.rvset from all sites, where $F_m$ is the largest fragment of $\mathcal{F}$. (2) It takes procedure evalDG $O(|G_d|)$ time to construct the dependency graph $G_d$, and to find whether $v_s$ reaches $v_{\text{true}}$ in $G_d$. Since $|G_d|$ is in $O(|V_f|^2)$, and $|V_f|$ is typically much smaller than $|F_m|$ in practice, the computational cost is bounded by $O(|F_m||V_f|)$. That is, the response time is also *independent* of the entire graph $G$.

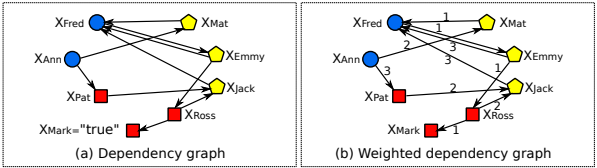

Figure 5: Dependency graphs

To check whether a pair of nodes connect in a fragment or in $G_d$, we use DFS/BFS search, and thus get the $O(|V_f||F_m|)$ (resp. $O(|V_f|^2)$) complexity. In fact any indexing techniques (*e.g.,* reachability matrix [31], 2-hop index [5]), parallel and graph partition strategies (*e.g.,* Pregel [21]) developed for *centralized graph query evaluation* can be applied here, which will lead to lower computational cost.

The analysis above completes the proof of Theorem 1.

**Remarks**. In theory, one can compute the transitive closure (TC) of a graph to decide whether a node can reach another. However, it is *impractical* to compute the TC over large graphs due to its time and space costs. Worse still, when the graphs are distributed, computing TC may incur *excessive unnecessary data shipments*. Indeed, we are not aware of any distributed algorithms that compute TC with performance guarantees on network traffic, even when indexing structures are employed (see [31] for a survey on such indexes). In contrast, we show that in the distributed setting, partial evaluation promises performance guarantees. Also observe that in practice, the size of $V_f$ is usually small [27].

## 4. DISTRIBUTED BOUNDED REACHABILITY QUERIES

We next develop a distributed evaluation algorithm for bounded reachability queries $q_{\text{br}}(s, t, l)$, to decide whether $\text{dist}(s, t) \leq l$. In contrast to reachability queries, to evaluate $q_{\text{br}}(s, t, l)$ we need to keep track of the distances for all pairs of nodes involved. Nevertheless, we show that the algorithm has the same performance guarantees as algorithm disReach.

**Theorem 2:** *Over a fragmentation $\mathcal{F} = (F, G_f)$ of a graph $G$, bounded reachability queries can be evaluated with the same performance guarantees as for reachability queries.* □

To prove Theorem 2, we outline an algorithm, denoted by disDist (not shown), for evaluating $q_{\text{br}}(s,t,l)$ over a fragmentation $\mathcal{F}$ of a graph $G$. It is similar to algorithm disReach for reachability queries (Fig. 3), but it needs different strategies for partial evaluation at individual sites and for assembling partial answers at the coordinator site. These are carried out by procedures localEval$_d$ and evalDG$_d$, respectively.

*Procedure* localEval$_d$. To evaluate $q_{\text{br}}(s,t,l)$, for each fragment $F_i$ and each in-node $v$ in $F_i$, we need to find $\text{dist}(v,t)$, the *distance* from $v$ to $t$. To do this, we find the *minimum* value of $\text{dist}(v,v') + \text{dist}(v',t)$ when $v'$ ranges over all virtual nodes in $F_i$ to which $v$ can reach. We associate a variable $X_{v'}$ with each such $v'$ to denote $\text{dist}(v',t)$ (*numeric value*). We express the partial answer for $v$ as a formula $v.\text{rf}$.

Procedure localEval$_d$ is similar to localEval, but differs in that for each virtual node $v$, if $v = t$, it assigns 0 to $v.\text{rf}$, and otherwise $v.\text{rf}$ is $X_v$. For each in-node $v \in \text{iset}$ and each virtual node $v' \in \text{oset}$, localEval$_d$ *locally* finds the distance from $v$ to $v'$ and uses a set st to collect formulas $v'.\text{rf} + \text{dist}(v,v')$ if $\text{dist}(v,v') < l$. The set $F_i$.rvset with *equations* $X_v = \min(v.\text{st})$ is sent to the coordinator site $S_c$.

*Procedure* evalDG$_d$. Given $F_i$.rvset from all the sites, procedure evalDG$_d$ assembles these partial answers to find the answer to $q_{\text{br}}(s,t,l)$ in $G$. As opposed to evalDG (Fig. 4), it builds an *edge weighted graph* $G_d = (V_d, E_d, L_d, W_d)$, where $(V_d, E_d, L_d)$ is a labeled dependency graph as defined before; and the *weight* $W_d(e)$ of $e$ is $\text{dist}(v_d, v'_d)$. Note that $|V_d| \leq |V_f|$ and $|E_d| \leq |V_f|^2$, where $G_f = (V_f, E_f)$ is the fragment graph of $\mathcal{F}$. The procedure then uses algorithm Dijkstra [32] to compute the distance $d$ from $X_s$ to $X_t$, in time $O(|E_d| + |V_d| \log |V_d|)$, where $X_s \in V_d$ denotes the node $s$ in $q_{\text{br}}(s,t,l)$. It returns true iff $d \leq l$. One can verify that $\text{dist}(s,t)$ in $G$ is equal to the distance from $X_s$ to $X_t$ in $G_d$.

**Example 5:** Given query $q_{\text{br}}(\text{Ann}, \text{Mark}, 6)$ posed on graph $G$ of Fig 1, disDist computes a set of equations of *arithmetic* formulas (not Boolean equations). The vectors for $F_2$ are:

| $F_i$ | $F_i.I$ | st | rvset |
|---|---|---|---|
| $F_2$ | Mat | $\{(x_{\text{Fred}} + 1)\}$ | $\{\{x_{\text{Mat}} = \min\{(x_{\text{Fred}} + 1)\},$ |
|  | Jack | $\{(x_{\text{Fred}} + 3)\}$ | $x_{\text{Jack}} = \min\{(x_{\text{Fred}} + 3)\}, x_{\text{Emmy}} =$ |
|  | Emmy | $\{(x_{\text{Fred}} + 3), (x_{\text{Ross}} + 1)\}$ | $\min\{(x_{\text{Fred}} + 3), (x_{\text{Ross}} + 1)\}\}$ |

After rvset is received by coordinator $\text{DC}_1$, procedure evalDG$_d$ first builds a weighted dependency graph $G_d$, shown in Fig 5(b). It then computes the shortest path from $X_{\text{Ann}}$ to $X_{\text{Mark}}$ by applying Dijkstra to $G_d$. It returns true since the length of the path is 6, satisfying the distance bound. □

One can verify that algorithm disDist (1) visits each site only once, (2) its total network traffic is bounded by $O(|V_f|^2)$, and (3) it takes at most $O(|F_m||V_f|)$ time, where $F_m$ is the largest fragment in $\mathcal{F}$. Moreover, indexing techniques [31] can be incorporated into localEval$_d$ and evalDG$_d$, to reduce the cost of local evaluation and hence, the response time (*e.g.,* with constant time via a distance matrix).

## 5. DISTRIBUTED REGULAR REACHABILITY QUERIES

We now develop techniques to distributively evaluate regular reachability queries. Given such a query $q_{\text{rr}}(s,t,R)$ and a fragmentation $\mathcal{F}$ of graph $G$, it is to find whether there exists a path $\rho$ from $s$ to $t$ in $G$ such that $\rho$ satisfies $R$. In contrast to (bounded) reachability queries, to evaluate $q_{\text{rr}}(s,t,R)$ we need to collect and transmit information about not only whether there are paths from a node to another, but also whether the paths satisfy the complex constraint imposed by $R$. The main result of this section is as follows.

**Theorem 3:** *On a fragmentation $\mathcal{F} = (F, G_f)$ of graph $G$, regular reachability queries $q_{\text{rr}}(s,t,R)$ can be evaluated (a) in $O(|F_m||R|^2 + |R|^2|V_f|^2)$ time, (b) by visiting each site once, and (c) with the total network traffic in $O(|R|^2|V_f|)^2)$, where $G_f = (V_f, E_f)$ and $F_m$ is the largest fragment in $F$.* □

To prove Theorem 3, we first introduce a notion of query automata (Section 5.1), and then present an evaluation algorithm based on query automata (Section 5.2).

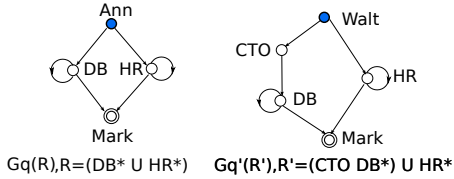

Figure 6: Query automaton $G_q(R)$

## 5.1 Query Automaton

To effectively check whether a path satisfies a regular expression $R$, we represent $R$ as a variation of nondeterministic finite state automata (NFA), referred to as query automaton.

A *query automaton* $G_q(R)$ of $q_{rr}(s, t, R)$ accepts paths $\rho$ that satisfy $R$. It is defined as $<V_q, E_q, L_q, u_s, u_t>$, where (1) $V_q$ is a set of states, (2) $E_q \subseteq V_q \times V_q$ is a set of transitions between the states, (3) $L_q$ is a function that assigns each state a label in $R$, and (4) $u_s$ and $u_t$ in $V_q$ are the start state and final state corresponding to $s$ and $t$, respectively. In contrast to traditional NFA, at state $u_v$, for each edge $(v, v')$ on a path, a transition $u_v \to u'_v$ can be made via $(u_v, u'_v) \in E_q$ if $L(v) = L_q(u_v)$ and $L(v') = L_q(u'_v)$. The automaton can be constructed in $O(|R|\log(|R|))$ time, using a conversion similar to that of [15]. It is of linear size in $|R|$.

We say that a state $u$ is a child of $u'$ (resp. $u'$ is a parent of $u$) if $(u', u) \in E_q$, i.e., $u'$ can transit to $u$.

**Example 6:** Recall $q_{rr}(\text{Ann}, \text{Mark}, R)$, the regular reachability query given in Example 1, where $R = (\text{DB}^* \cup \text{HR}^*)$. Its query automaton $G_q(R)$ is depicted in Fig 6. The set $V_q$ has four states Ann, DB, HR, Mark, where the start and final states are Ann and Mark, respectively. The set $E_q$ of transitions is {(Ann,DB), (DB,DB), (DB,Mark), (Ann,HR), (HR,HR), (HR,Mark)}. In contrast to NFA, it is to accept paths in, e.g., $G$ of Fig. 1, and its transitions are made by matching the labels of its states with the job labels on the paths (except the start and final states, which are labeled with name).

As another example, consider query $q_{rr}(\text{Walt}, \text{Mark}, R')$, where $R' = ((\text{CTO DB}^*) \cup \text{HR}^*)$. Figure 6 shows its query automaton, which has 5 states and 7 transitions, with Walt and Mark as its start state and final state, respectively. □

We say that a node $v$ in $G$ is a *match* of a state $u_v$ in $G_q(R)$ iff (1) $L(v) = L_q(u_v)$, and (2) there exist a path $\rho$ from $v$ to $t$ and a path $\rho'$ from $u_v$ to $u_t$, such that $\rho$ and $\rho'$ have the same label. The lemma below shows the connection between $q_{rr}(s, t, R)$ and $G_q(R)$, which is easy to verify.

**Lemma 4:** *Given a graph $G$, $q_{rr}(s, t, R)$ over $G$ is true if and only if $s$ is a match of $u_s$ in $G_q(R)$.* □

## 5.2 Distributed Query Evaluation Algorithm

We next present an algorithm to evaluate regular reachability queries over a fragmentation $\mathcal{F}$ of a graph $G$. The algorithm, denoted as disRPQ (not shown), evaluates $q_{rr}(s, t, R)$ based on partial evaluation in three steps, as follows.

(1) It first constructs the query automaton $G_q(R)$ of $q_{rr}(s, t, R)$ at site $S_c$, and posts $G_q$ to each fragment in $\mathcal{F}$.

(2) Upon receiving $G_q(R)$, each site invokes procedure localEval$_r$ to compute a *partial answer* to $q_{rr}(s, t, R)$ by using $G_q$, *in parallel*. The partial answer at each fragment $F_i$, denoted as $F_i$.rvset, is a set of *vectors*. Each entry in a vector is a Boolean formula (as will be discussed shortly).

(3) The partial answer is sent back to the coordinator site $S_c$. The site $S_c$ collects $F_i$.rvset from each site and assembles

---

**Procedure** localEval$_r$ /* executed locally at each site, in parallel */
*Input:* A fragment $F_i$, a query automaton $G_q(V_q, E_q, L_q, u_s, u_t)$.
*Output:* Partial answer to $q_{rr}$ in $F_i$ (a set rvset of vectors).
1. $F_i$.rvset := $\emptyset$; iset:= $F_i.I$; oset:= $F_i.O$;
2. **if** $s \in F_i$ **then** iset:= iset $\cup \{s\}$    /* $s$ denoted by $u_s$ */
3. **if** $t \in F_i$ **then** oset:= oset $\cup \{t\}$;  /* $t$ denoted by $u_t$ */
4. **for each** node $v \in V_i \setminus$ oset **do** $v$.visit := false;
5. **for each** node $v \in$ oset **do**
6.    $v$.rvset := $\emptyset$;
7.    **for each** node $u \in V_q$ **do**
8.       **if** $v = t$ **and** $u = u_t$ **then** $v$.rvec[$u_t$] := true;
9.       **else if** $L(v) = L_q(u)$ **then** $v$.rvec[$u$] := $X_{(v,u)}$;
10.      **else** $v$.rvec[$u$] := false;
11.   $v$.visit := true;
12. **for each** node $v \in$ iset **do**
13.    $v$.rvec := cmpRvec($v, F_i, q_{rr}, G_q$);
14.    $F_i$.rvset := $F_i$.rvset $\cup$ $v$.rvec;
15. send $F_i$.rvset to the coordinator site $S_c$;

**Procedure** cmpRvec
*Input:* A node $v$, a fragment $F_i$, and
    a query automaton $G_q(V_q, E_q, L_q, u_s, u_t)$.
*Output:* The vector $v$.rvec of $v$, consisting of Boolean formulas.
1. **if** $v$.visit = true **then return** $v$.rvec;
2. **for each** node $v_q \in V_q$ **do** rvec[$v_q$] := false;
3. **for each** node $w \in C(v, F_i)$ **do**
4.    **if** $w$.visit = false **then**
5.       $w$.rvec := cmpRvec($w, F_i, q_{rr}, G_q(R)$);
6.    **for each** node $v_q \in V_q$ **do**
7.       **if** $L(v) = L_q(v_q)$ **then**
8.          rvec[$v_q$] := rvec[$v_q$] $\vee$ cmposeVec($v_q, w, w$.rvec, $G_q(R)$);
9. $v$.visit := true;
10. **return** rvec;

**Figure 7: Procedure localEval$_r$ and cmpRvec**

them into a set RVset of vectors of Boolean formulas. It then invokes procedure evalDG$_r$ to solve these equations and find the final answer to $q_{rr}(s, t, R)$ in $G$.

We now present procedures localEval$_r$ and evalDG$_r$.

**Local evaluation**. We first formulate the partial answer $v$.rvec at each node $v$ in a fragment $F_i$. It indicates whether $v$ is a match of some state $u$ in the query automaton $G_q$, i.e., $v$ reaches $t$ *and moreover*, satisfies the constraints imposed by $G_q$ (Lemma 4). Hence we define $v$.rvec to be a *vector* of $O(|V_q|)$ entries, where $V_q$ is the set of states in $G_q$. For each state $u$ in $V_q$, the entry $v$.rvec[$u$] is a *Boolean formula* indicating whether node $v$ matches state $u$. In contrast to its counterparts for (bounded) reachability queries, here $v$.rvec is a *vector* of Boolean formulas, instead of a single formula.

Observe that $v$ matches a state $u_v$ if and only if (1) $L(v) = L(u_v)$, and (2) either $v$ is $t$, or there exists a child $w$ of $v$ and a child $u_w$ of $u_v$ such that $w$ matches $u_w$. To cope with virtual nodes, for each $w \in F_i.O$ and each state $u_w \in V_q$, we introduce a Boolean variable $X_{(w, u_w)}$, denoting whether $w$ matches $u_w$. The vector of each in-node $v$ in $F_i.I$ consists of formulas defined in terms of these Boolean variables.

Based on these, we give procedure localEval$_r$ in Fig. 7. It first initializes a set $F_i$.rvset of vectors, and puts the in-nodes $F_i.I$ and virtual nodes $F_i.O$ of $F_i$ in sets iset and oset, respectively (line 1). If $s$ (resp. $t$) is in $F_i$, localEval includes $s$ (resp. $t$) in iset (resp. oset) as well (lines 2-3). For each node $v$ in $F_i$, it associates a *flag* $v$.visit to indicate whether $v$.rvec is already computed, and initializes it to be false if $v$ is not in oset (line 4). It then initializes the vector $v$.rvec for each virtual node $v$ of $F_i$ (lines 5-11), as follows. If $v = t$, then $v$.rvec[$u_t$] is assigned true (line 8). Otherwise

1310

for each state $u$ in $G_q$, if $u$ and $v$ have the same label, then $v$.rvec$[u_t]$ is a *Boolean variable* $X_{(v,u)}$, indicating whether $v$ *matches* $u$ (line 9); if not, $v$.rvec$[u]$ is false (line 10). Since $v$.rvec is initialized (lines 6-10), localEval sets $v$.visit to be true (line 11). Then for each in-node $v$, localEval$_r$ invokes procedure cmpRvec to partially compute the vector of $v$, and extends $F_i$.rvset with $v$.rvec (lines 12-14). After all in-nodes are processed, $F$.rvset is sent to site $S_c$ (line 15).

Procedure cmpRvec computes the vector $v$.rvec for a node $v$, as follows. If $v$.visit is true, it returns $v$.rvec (line 1). Otherwise, it initializes a vector rvec (lines 2). The procedure then computes $v$.rvec following Lemma 4. For each child $w$ of $v$, if $w$ is not visited, then $w$.rvec is computed via a recursive call of cmpRvec (lines 3-5; here $C(v, F_i)$ denotes the set of children of $v$ in $F_i$). After $w$.rvec is known, for each state $v_q$ in $G_d$, cmpRvec checks if $v$ and $v_q$ have the same label (lines 6-7); if so, it uses $w$.rvec$[v'_q]$ to compute rvec$[v_q]$ via procedure cmposeVec (line 8). After $v$.rvec$[v_q]$ is computed, $v$.visit is set true (line 9) and $v$.rvec$[v_q]$ is returned (line 10).

Procedure cmposeVec (not shown) takes a state $v_q$ and a node $w$ as input, and constructs a formula $f$ using formulas in $w$.rvec. Initially $f$ is false. For each child state $v'_q$ of $v_q$, it checks whether $w$ and $v'_q$ have the same label. If so, $f$ is extended by taking $w$.rvec$[v'_q]$ as a disjunct. The formula $f$ is returned after all child states of $v_q$ is processed.

**Example 7:** Given q$_{rr}$(Ann, Mark, $R$), the query of Example 1 posed on the distributed graph $G$ of Fig. 1, procedure localEval$_r$ evaluates the query on $F_2$ as follows. For each virtual node of $F_2$, it initializes its vector, *e.g.*, the vector of Ross is (false, false, $X_{(Ross,HR)}$, false), corresponding to the states (Ann, DB, Mark, HR) in query automaton $G_q(R)$ (see Fig. 6). It then invokes procedure cmpRvec to compute the vector of each in-node of $F_2$. For instance, consider in-node Emmy. Since (1) Emmy is an HR that matches state HR in $G_q$, and (2) Emmy has a child Ross that may match state HR, the formula Emmy.[HR] is extended to $X_{(Ross,HR)}$ by procedure cmposeVec. The final vectors for $F_2$ are:

| fragment | in-node | rvec(Ann, DB, HR, Mark) | | | |
|---|---|---|---|---|---|
| $F_2$ | Mat | false | false | $X_{(Fred,HR)}$ | false |
| | Jack | false | false | false | false |
| | Emmy | false | false | $X_{(Ross,HR)}$ | false |

□

**Assembling**. Procedure evalDG$_r$ (not shown) collects the partial answers from all the sites into a set RVset, and assemble them to compute the answer to q$_{rr}(s,t,R)$ at the coordinator site $S_c$. It is similar to procedure evalDG given in Fig. 4, except that it uses a different notion of dependency graphs. Here the *dependency graph* $G_d$ of RVset is defined as $(V_d, E_d, L_d)$, where (a) for each in-node $v$ and each entry $u$ of its vector $v$.rvec in RVset, there is a node $v_{d(v,u)} \in V_d$, (b) $L_d(v_{d(v,u)}) = v$.rvec$[u]$, a formula of the form $\bigvee X_{(v',u')}$; and (c) there is an edge $(v_{d(v,u)}, v_{d(v',u')}) \in E_d$ if and only if $X_{(v',u')}$ appears in $L_d(v_{d(v,u)})$. In other words, the node set $V_d$ of $G_d$ is defined in terms of both in-nodes in the fragments of $\mathcal{F}$ and the states in the query automaton $G_q$.

Procedure evalDG$_r$ constructs the dependency graph $G_d$ of RVset, and checks whether $v_d(s, u_s)$ can reach $v_d(u, u')$ for some node $u$, where $L_d(v_{u,u'})$ is true. One can verify that $s$ matches $u_s$ iff there exists a node $v_{d(u,u')} \in V_d$ with $L_d(v_{u,u'}) =$ true, and $v_d(s, u_s)$ reaches $v_{d(u,u')}$.

**Example 8:** Consider again query q$_{rr}$(Ann, Mark, $R$) posed on the graph $G$ of Fig. 1. The vector sets $F_i$.rvset are computed in parallel in all fragments $F_i$, as described in Ex-ample 7. Upon receiving $F_i$.rvset from all the sites, procedure evalDG$_r$ first builds a dependency graph $G_d$ based on the vector sets, as shown in Fig 8. Each node, *e.g.*, $v_d$(Ann, Ann) is shown together with its label, *e.g.*, $X_{(Mat,HR)}$. It then checks whether node $v_d$(Ann, Ann) reaches a node with label true, which is node $v_d$(Ross, HR) here. It returns true as the query answer, as there is a path (Ann, Mat, Fred, Emmy, Ross, Mark) satisfying the regular expression $R$. □

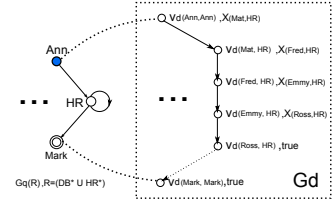

**Figure 8:** Assembling with dependency graph

**Correctness and complexity**. One can readily verify the following. (1) The algorithm disRPQ always terminates. (2) Given a query q$_{rr}(s, t, R)$ and a fragmentation $\mathcal{F}$ of graph $G$, algorithm disRPQ returns true iff there exists a path $\rho$ from $s$ to $t$ in $G$ such that $\rho$ satisfies $R$. To complete the proof of Theorem 3, observe the following about its complexity.

*The number of visits*. Each site is visited only once, when the query automaton is posted by the coordinator site.

*Total network traffic*. The communication cost includes the following: (1) $O(|G_q|\mathsf{card}(F))$ for sending query automaton $G_q(R)$ to each site, where card$(F)$ is the number of fragments, and $|G_q|$ is in $O(|R|)$; and (2) $O(|R|^2|F_i.I||F_i.O|)$ for sending partial answers from each fragment $F_i$ to the coordinator site. Putting these together, the total network traffic is in $O(|R|^2|V_f|^2)$, where $V_f$ is the total number of virtual nodes, since the number card$(F)$ of fragments and query size $|R|$ are much smaller than $|V_f|$ in practice. Note that the communication cost is *independent of* the entire graph $G$.

*Total computation*. It takes $O(|R|^2 * |F_m|)$ time to compute the vector set in each fragment, *in parallel*, where $|F_m|$ is the size of the largest fragment $F_m$ in $\mathcal{F}$. To see this, observe that at each node $v$, it takes at most $O(|C(v, F_m)| * |R|^2)$ time to construct its vector, for each child of $v$ in $C(v, F_m)$. Moreover, each node is visited once and its vector is computed once. Thus, in total it takes at most $O(|F_m||R|^2)$ time to compute all the vectors. The assembling phase takes up to $O(|R|^2|V_q|)^2)$ time. Taking these together, the total computation time is in $O(|F_m||R|^2 + |R|^2|V_f|)^2)$.

## 6. DISTRIBUTED REACHABILITY WITH MAPREDUCE

We next present a simple MapReduce algorithm to evaluate regular reachability queries. This algorithm just aims to demonstrate how easy to support our techniques in the MapReduce framework. More advanced MapReduce algorithms can be readily developed based on partial evaluation.

MapReduce [7] is a software framework to support distributed computing on large datasets with a large number of computers (nodes). (1) The data are partitioned into a collection of key/value pairs. Each pair is assigned to a node (*mapper*) identified by its key. (2) Each mapper processes its key/value pairs, and generates a set of intermediate key/value pairs, by using a Map *function*. These pairs are hash-partitioned based on the key. Each partition is sent to a node (*reducer*) identified by the key. (3) Each reducer

1311

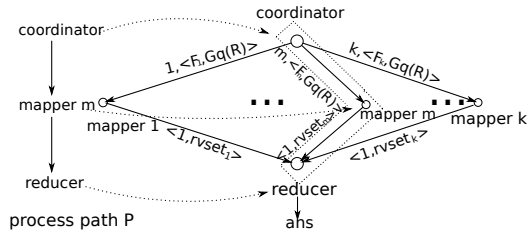

**Figure 9: Processing path of algorithm reduceRPQ**

produces key/value pairs via a Reduce *function*, and writes them to a distributed file system as the result [7].

Our MapReduce algorithm, MRdRPQ, is illustrated in Fig. 9 and given in Fig. 10. It evaluates $q_{rr}(s, t, R)$ on graph $G$ using procedures preMRPQ, mapRPQ and reduceRPQ. We next present the three procedures in details.

*Procedure* preMRPQ. A coordinator first generates the query automaton $G_q(R)$ of $q_{rr}(s, t, R)$ (line 1; see Section 5). The graph $G$ is then partitioned into $K$ fragments (line 2) using some strategy parG, where $K$ is the number of mappers. Each fragment $F_i$ is represented as a key/value pair, where the key is $i \in [1, K]$, and its value is a pair $<F_i, G_q(R)>$ (lines 3-4). It is sent to mapper $M_i$ along with $G_q(R)$ (line 5).

Graph partitioning is conducted implicitly by MapReduce implementation (*e.g.,* Hadoop), provided the number $K$ of mappers and the average size $\lceil \frac{|G|}{K} \rceil$ of fragments (line 2). To explore the maximum parallelism we want the fragments to be of equal size; hence $\lceil \frac{|G|}{K} \rceil$. One may also want to minimize $\sum_{F_i \in F} |F_i.I||F_i.O|$, where $F_i.I$ (resp. $F_i.O$) is the set of in-nodes (resp. virtual nodes) of fragment $F_i$. However, this partition problem is intractable [10]. In our implementation we used Hadoop's default partitioning strategy.

*Procedure* mapRPQ *at each mapper*. Upon receiving a pair $<i, (F_i, G_q(R))>$, procedure mapRPQ is triggered at mapper $M_i$, *in parallel*. It simply uses procedure localEval$_r$ of Fig. 7 as its Map function, and computes a key/value pair $<1, \text{rvset}_i>$ (line 1), where rvset$_i$ is the vector set as described in Section 5. It sends the pair to a reducer $R_o$. Note that pairs from all the mappers are sent to the same reducer.

*Procedure* reduceRPQ *at the reducer* $R_o$. After collecting the key/value pairs from all the mappers, the reducer puts these pairs in a set RVset (lines 1-3). It then invokes the assembling procedure evalDG$_d$ (see Section 5) as the Reduce *function* to compute the answer ans to $q_{rr}$ in $G$ (line 4), and writes a pair $<0, \text{ans}>$ to the distributed file system (line 5).

**Correctness and complexity.** The correctness of algorithm MRdRPQ immediately follows from the correctness of algorithm disRPQ (see Section 5). Following [1], we analyze the performance of MRdRPQ using the *elapsed communication cost* ECC (data volume cost), which measures the total time cost of (parallel) data shipment. We define a *process path* $P$ of MRdRPQ to be a path from the coordinator to the reducer, passing a single mapper (see Fig. 9). The cost of a process path $\alpha$ is the sum of the *size of input data* shipped to the nodes on $\alpha$, following an edge of $\alpha$. The ECC of MRdRPQ is the maximum cost over all process paths.

The ECC analysis unifies the time and network traffic costs of a MapReduce algorithm. It does not count the in-memory computation cost of the Map and Reduce functions. Nevertheless, (1) any indexes and compression techniques developed for centralized graph query evaluation can be adopted

---

**Procedure** preMRPQ
*Input:* Graph $G$, regular reachability query $q_{rr}(s, t, R)$, integer $K$.
*Output:* Lists of key/value pairs to be sent to mappers.

1. construct query automaton $G_q(R)$;/*executed at coordinator*/
2. glist := parG$(G, K, \lceil \frac{|G|}{K} \rceil)$; /* graph partition */
3. **for each** fragment $F_i \in$ glist $(i \in [1, K])$ **do**
4.     pair $L := <i, (F_i, G_q(R))>$;
5.     send $L$ and $G_q(R)$ to mapper $i$;

**Procedure** mapRPQ   /* executed at each mapper */
*Input:* A key/value pair $L = <i, (F_i, G_q(R))>$.
*Output:* A key/value pair rdpair.

1. rvset$_i$ := localEval$_r(F_i, G_q(R))$;
2. send localEval$_r(F_i, G_q(R))$ to a reducer;

**Procedure** reduceRPQ   /* executed at a single reducer */
*Input:* A list of key/value pairs.
*Output:* The Boolean value ans to $q_{rr}$ in $G$.

1. set RVset := $\emptyset$;
2. **for each** pair $<1, \text{rvset}_i>$ in rdlist **do**
3.     RVset:= RVset $\cup$ rvset$_i$;
4. ans:= evalDG$_r$(RVset);
5. **return** $<0, \text{ans}>$;

**Figure 10: Algorithm MRdRPQ**

by mappers, as remarked earlier, (2) further MapReduce steps can be used to implement both Map and Reduce functions, and (3) network traffic dominates the total computation time for real-life large graphs [1].

For algorithm MRdRPQ, one can verify the following. (1) The input size of each mapper is bounded by $O(|F_m|)$, where $F_m$ is the largest fragment returned by parG. (2) The input size of the reducer is bounded by $O(|R|^2|V_f|^2)$, where $V_f$ is the set of nodes in the fragment graph $G_f$. Putting these together, the ECC of mapRPQ is $O(|F_m| + |R|^2|V_f|^2)$.

## 7. EXPERIMENTAL EVALUATION

We next present an experimental study of our distributed algorithms. Using real-life and synthetic data, we conducted four sets of experiments to evaluate the efficiency and communication costs of algorithms disReach (Section 3), disDist (Section 4), disRPQ (Section 5) and the MapReduce algorithm MRdRPQ (Section 6) on Amazon EC2.

**Experimental setting.** We used the following data.

*(1) Real-life graphs.* For (bounded) reachability queries, we used the following[4]: (a) a social network LiveJournal, (b) a communication network WikiTalk, (c) two Web graphs BerkStan and NotreDame, and (d) a product co-purchasing network Amazon. The sizes of these graphs are shown below.

| dataset | $|V|$ | $|E|$ |
|---|---|---|
| LiveJournal | 2,541,032 | 20,000,001 |
| WikiTalk | 2,394,385 | 5,021,410 |
| BerkStan | 685,230 | 7,600,595 |
| NotreDame | 325,729 | 1,497,134 |
| Amazon | 262,111 | 1,234,877 |

For regular reachability queries, we used the following graphs with attributes on the nodes: (a) Citation[5], in which nodes represent papers with id and venue, and edges denote citations, (b) MEME[5], a blog network in which nodes are Web pages and edges are links, (c) Youtube[6], a social network in which each node is a video with attributes (*e.g.,* category), and each edge indicates a recommendation, and (d) Internet[7], where each node is a system labeled with its id and location, and each edge represents internet connection. The datasets are summarized below, where $|L|$ is the size of node label set, and card$(F)$ is the number of the fragments generated for regular reachability queries (see below).

---

[4] *http://snap.stanford.edu/data/index.html*
[5] *http://www.arnetminer.org/citation/*
[6] *http://netsg.cs.sfu.ca/youtubedata/*

| dataset | $|V|$ | $|E|$ | $|L|$ | card$(F)$ |
|---|---|---|---|---|
| Citation | 1,572,278 | 2,084,019 | 6300 | 10 |
| MEME | 700,000 | 800,000 | 61065 | 11 |
| Youtube | 234,452 | 454,942 | 12 | 12 |
| Internet | 57,971 | 103,485 | 256 | 10 |

*(2) Synthetic data.* We designed a generator to produce large graphs, controlled by the number $|V|$ of nodes, the number $|E|$ of edges, and the size $|L|$ of node labels.

*(3) Graph fragmentation.* We randomly partitioned real-life and synthetic graphs $G$ into a set $F$ of fragments, controlled by card$(F)$ and the average size of the fragments in $F$ (the sum of the numbers of nodes and edges), denoted by size$(F)$. Unless stated otherwise, size$(F) = |G|/$card$(F)$.

*(4) Query generator.* We randomly generated (a) reachability queries, (b) bounded reachability queries with bound $l$, and (c) regular reachability queries from a set $L$ of labels.

*(5) Algorithms.* We implemented the following algorithms in Java: (A) disReach, disReach$_n$ and disReach$_m$ for reachability queries, where (a) disReach$_n$ ships all the fragments to a coordinator in parallel, which calls a centralized BFS algorithm to evaluate the query [31]; and (b) disReach$_m$, a message-passing based distributed BFS algorithm following [21] (see details below); (B) disDist and disDist$_n$ for bounded reachability queries, where disDist$_n$ is similar to disReach$_n$; (C) disRPQ, disRPQ$_n$ and disRPQ$_d$ for regular reachability queries, where disRPQ$_n$ is similar to disReach$_n$, and disRPQ$_d$ is a variant of the algorithm of [30] (see Section 1); and (D) the MapReduce algorithm MRdRPQ.

Following [21], algorithm disReach$_m$ assigns a worker $S_i$ for each fragment $F_i$, and a master $S_c$ that maintains the fragment graph (see Section 2). (i) Each node $v$ in the fragments has a status $l(v) \in$ {inactive, active}, initially inactive. (ii) A *message* "T" can be sent only from *active* nodes $v_1$ (*i.e.*, $l(v_1) = $ active) to their *inactive children* $v_2$ (*i.e.*, $l(v_2) = $ inactive), which then become *active*. (iii) no *active* node can become *inactive* again. (iv) $S_i$ can send "T", "idle", or a *virtual node* of $F_i$ as a message to $S_c$.

Upon receiving a reachability query q$_r(s,t)$, $S_c$ posts q$_r$ to all the workers $S_i$. For the fragment $F_i$ that contains the node $s$ specified in q$_r(s,t)$, its worker $S_i$ changes $l(s)$ to active, and sends a message "T" to its immediate inactive children, which in turn propagate "T" following a BFS traversal to inactive nodes. During the propagation, (i) if "T" reaches an inactive virtual node $v$, $S_i$ sends a message $v$ to $S_c$, which redirects the message to workers $S_j$ where the fragments $F_j$ has inactive in-node $v$; $S_j$ then makes $v$ active, and propagates "T" along the same lines in $F_j$; (ii) if "T" reaches the node $t$ in q$_r(s,t)$, $S_i$ sends message "T" to $S_c$, and algorithm disReach$_m$ returns true, indicating that q$_r(s,t) = $ true; and (iii) when no message is propagating in $S_i$, it sends message "idle" to $S_c$. Algorithm disReach$_m$ returns false if all the workers send "idle" to it.

*Machines.* We deployed these algorithms on Amazon EC2 High-Memory Double Extra Large instances[8]. Each site

[7]http://www.caida.org/data/
[8]http://aws.amazon.com/ec2/

| Datasets | Time(second) | | | Traffic(MB) | | |
|---|---|---|---|---|---|---|
| | disReach | disReach$_n$ | disReach$_m$ | disReach | disReach$_n$ | disReach$_m$ |
| LiveJournal | 12.03 | 27.52 | 186.55 | 174 | 1800 | 27 |
| WikiTalk | 3.32 | 9.95 | 41.42 | 80 | 726 | 19 |
| BerkStan | 3.25 | 8.51 | 40.31 | 29 | 555 | 11 |
| NotreDame | 0.83 | 3.77 | 13.32 | 14 | 147 | 7 |
| Amazon | 0.55 | 2.55 | 7.86 | 10 | 120 | 5 |

**Table 2: Efficiency and data shipment: real life data**

stored a fragment. Each experiment was run 5 times and the average is reported here.

**Experimental results.** We next present our findings.

**Exp-1: Efficiency and scalability of disReach.**

*Efficiency.* We first evaluated the efficiency of disReach, disReach$_n$ and disReach$_m$. Fixing card$(F) = 4$, we randomly generated 100 reachability queries (where around 30% return "true"), and report the average evaluation time and the network traffic in Table 2. The results show that disReach is far more efficient than disReach$_n$ and disReach$_m$. For example, on Amazon, disReach takes only 20% of the running time of disReach$_n$, and 6% of that of disReach$_m$. On the real datasets it takes 4 seconds in average.

For the network traffic of disReach$_m$, we counted the total number of messages sent between the workers and the master. Table 2 shows that in average, the network traffic of disReach is only 9% of that of disReach$_n$ (*i.e.*, the size of the original graphs), but is not as good as that of disReach$_m$. Indeed, the data shipment of disReach$_m$ is linear in the number of the total virtual nodes. However, this reduction comes at the cost of serializing operations that can be conducted in parallel, as indicated by its extra running time (Table 2). Moreover, it has no bound on the number of visits to each site; for instance, when card$(F) = 4$ on Amazon, the four sites were visited about 2500 times in total.

*Scalability.* To evaluate the scalability with card$(F)$, we used LiveJournal as the dataset and varied card$(F)$ from 2 to 20. We used the same set of queries as above. Fig. 11(a) shows that the larger card$(F)$ is, the less time disReach and disReach$_n$ take. For disReach, this is because *partial evaluation* of localEval takes less time on smaller fragments. For disReach$_n$, while the evaluation time on the restored graph remains stable (about 10 seconds), it takes less time to ship each fragment to the coordinator when card$(F)$ increases. In contrast, the larger card$(F)$ is, the more costly disReach$_m$ is. Indeed, smaller fragments require more frequent visits and thus, more communication cost.

To evaluate the scalability with the average size$(F)$ of fragments, we generated synthetic graphs following the *densification law* [20], by fixing card$(F) = 8$ and varying the size of the graphs from 280K to 2.52M. As shown in Fig. 11(b), when size$(F)$ is increased, so is the running time of all these algorithms, as expected. Nonetheless, disReach scales well with size$(F)$, and is less sensitive to size$(F)$ than the others.

We also tested disReach and disReach$_m$ over a larger synthetic graph, which has 36M nodes and 360M edges. We varied card$(F)$ from 10 to 20 in 2 increments. The results, shown in Fig 11(c), tell us the following. (1) disReach scales well with card$(F)$, and takes less time over larger card$(F)$, and (2) disReach$_m$ takes more time when card$(F)$ gets larger. The results are consistent with the observation of Fig 11(a).

**Exp-2: Efficiency of disDist.** This set of experiments evaluated the performance of disDist and disDist$_n$. Using WikiTalk, we varied card$(F)$ from 2 to 20, and ran-

1313

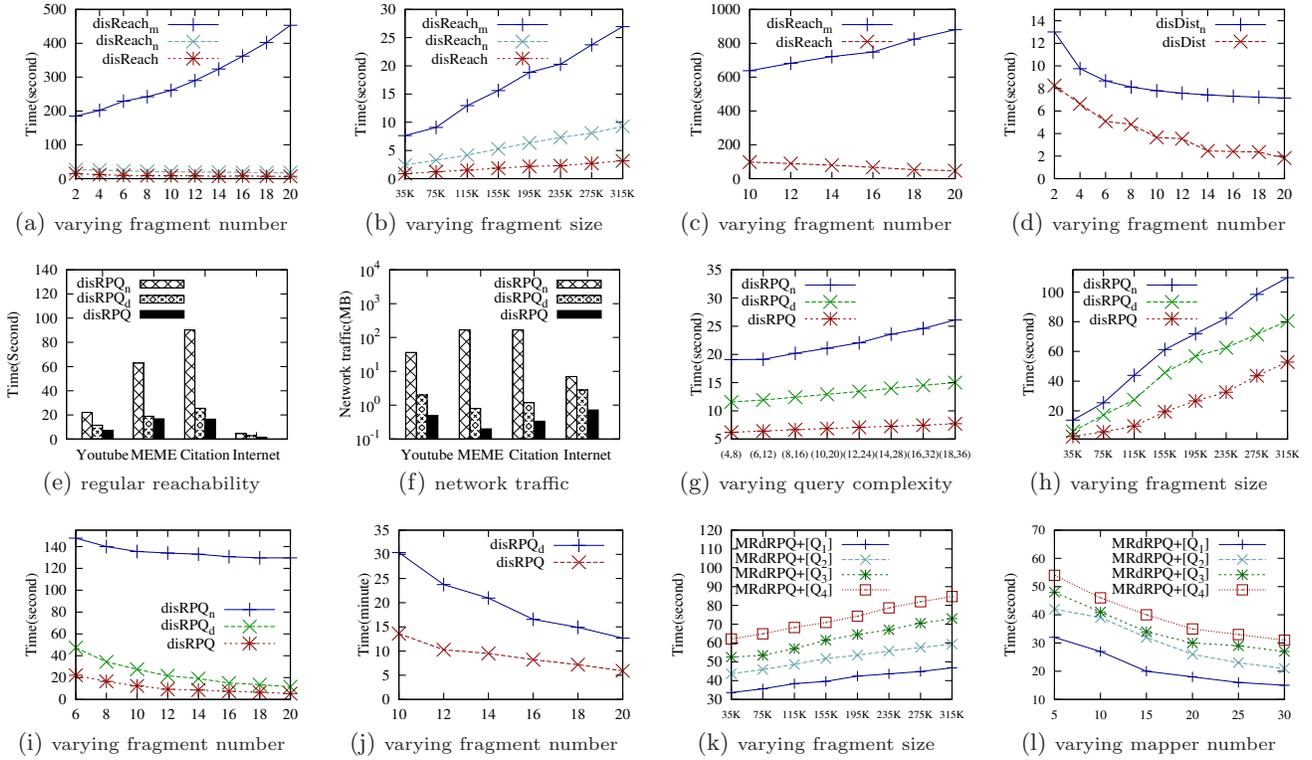

**Figure 11: Performance evaluation**

domly generated 100 bounded reachability queries with $l=10$. Fig. 11(d) shows that (1) disDist outperforms $\text{disDist}_n$ by 62.5% in average, and (2) disDist and $\text{disDist}_n$ take less time over larger $\text{card}(F)$, for the same reason as given above.

The performance of disDist and $\text{disDist}_n$ (not shown) are consistent with their counterparts (disReach and $\text{disReach}_n$).

**Exp-3: Efficiency and scalability of disRPQ.**

*Efficiency.* The third set of experiments focused on the performance of algorithms disRPQ, $\text{disRPQ}_n$ and $\text{disRPQ}_d$ [30], for regular reachability queries. We specify the complexity of such a query in terms of $(|V_q|, |E_q|, |L_q|)$, where $V_q$, $E_q$ and $L_q$ are the sets of states, transitions and node labels in its query automaton, respectively (see Section 5.1).

We first evaluated the response time and network traffic of these algorithms on the four real-life datasets described earlier, with $|V|, |E|, |L|$ and $\text{card}(F)$ given there. We generated 30 regular reachability queries with $(|V_q|=8, |E_q|=16, |L_q|=8)$, and report their average time (resp. network traffic) in Fig. 11(e) (resp. Fig 11(f)). We find the following: (1) disRPQ is more efficient than $\text{disRPQ}_n$ and $\text{disRPQ}_d$; indeed, the running time of disRPQ is 61.8%, 88%, 64.8% and 56.6% of that of $\text{disRPQ}_d$ on Youtube, MEME, Citation and Internet, respectively; and (2) disRPQ incurs less network traffic than the other algorithms: at most 25% of data shipped by $\text{disRPQ}_d$ and 3% of that of $\text{disRPQ}_n$ in average.

To evaluate the impact of query complexity, we used Youtube and generated 40 regular reachability queries by varying $|V_q|$ from 4 to 18 and $|E_q|$ from 8 to 36, while fixing $|L_q|=8$. Fig. 11(g) shows that (1) all the algorithms take longer to answer larger queries, and (2) disRPQ and $\text{disRPQ}_d$ are less sensitive to the size of queries than $\text{disRPQ}_n$.

*Scalability.* We generated synthetic graphs by fixing $\text{card}(F) = 10$ while varying the size of the graphs from 350K to 3.15M. We tested 30 queries with $|V_q|=8$, $|E_q|=16$ and $|L_q|=8$, and report the average running time in Fig. 11(h). The result shows that disRPQ scales well with $\text{size}(F)$, and performs better than $\text{disRPQ}_d$ and $\text{disRPQ}_n$. Moreover, it is efficient: disRPQ takes 16 seconds on graphs with 1.5M (million) nodes and 2.1M edges. In addition, the larger $\text{size}(F)$ is, the longer the three algorithms take, as expected.

To evaluate the scalability $\text{card}(F)$, we generated graphs with 1.2M nodes and 4.8M edges, and varied $\text{card}(F)$ from 6 to 20. As shown in Fig. 11(i), the larger $\text{card}(F)$ is, the less time disRPQ takes, since it conducts partial evaluation on smaller fragments by exploring parallel computation. This confirms our complexity analysis for disRPQ (Section 5). Indeed, the time taken by disRPQ when $\text{card}(F)=6$ is reduced by 75% when $\text{card}(F)=20$. Similarly, $\text{disRPQ}_d$ and $\text{disRPQ}_n$ take less time when $\text{card}(F)$ is increased.

In addition, we evaluated the scalability of disRPQ and $\text{disRPQ}_d$ over large synthetic graphs. Fixing $|V|=36M$, $|E|=360M$ and $|L|=50$, we varied $\text{card}(F)$ from 10 to 20 in 2 increments. As shown in Fig 11(j), (1) both algorithms scale well with $\text{card}(F)$, and take less time when $\text{card}(F)$ increases; and (2) disRPQ consistently outperforms $\text{disRPQ}_d$.

**Exp-4: Efficiency of MRdRPQ.** Finally, we evaluated the efficiency and scalability of MRdRPQ, implemented using Hadoop (*http://hadoop.apache.org*), and deployed on Amazon EC2, where each instance serves as a mapper. We use Youtube and four sets of $\text{q}_{rr}$ $Q_1$, $Q_2$, $Q_3$, $Q_4$ of different complexities (4, 6, 8), (6, 8, 8), (10, 12, 8), (12, 14, 8), respectively.

To evaluate the scalability of MRdRPQ, we fixed the number of mappers as 10, and varied the graph size from 350K to 3.15M. As shown in Fig. 11(k), MRdRPQ scales well with $\text{size}(F)$. Moreover, the larger $\text{size}(F)$ is or the more complex a query is, the longer time MRdRPQ takes, as expected. To evaluate its scalability with the number $|M|$ of mappers, we varied $|M|$ from 5 to 30. As shown in Fig. 11(l), it takes less time of MRdRPQ to evaluate queries with more mappers. Indeed, the time taken by MRdRPQ using 5 mappers is reduced by 50% when 30 mappers are used for $Q_1$.

We also find that disRPQ takes 17.4% of the running time of MRdRPQ and 3.7% of its network traffic on Youtube. The extra cost of MRdRPQ is incurred in the Map phase of the MapReduce framework, for distributing data to mappers.

**Summary.** From the experimental results we find the following. (1) All of our algorithms scale well with the size of graphs, the number of fragments, and the complexity of queries (for disRPQ and MRdRPQ). (2) Our algorithms are efficient even on *randomly* partitioned graphs. For instance, (a) disReach takes 20% and 6% of the running time of disReach$_n$ and disReach$_m$ over Amazon, and takes in average 4 seconds over all real life datasets; and (b) disRPQ takes 67.8% and 46% of the time of disRPQ$_d$ [30], and ships 47.9% and 45.9% of the data sent by disRPQ$_d$, on real-life and synthetic graphs in average, respectively. Overall our algorithms ship no more than 11% of the entire graphs in average. (3) Partial evaluation works well in the MapReduce model, as verified by the performance of MRdRPQ.

## 8. CONCLUSION

We have provided algorithms for evaluating a group of reachability queries on distributed graphs based on partial evaluation, possess performance guarantees on *the number of visits* to each site, the total *network traffic*, and on the *response time*. Moreover, they are *generic*: no constraints is posed on how the graphs are partitioned and distributed. We have also shown that partial evaluation can be naturally conducted as MapReduce. Our experimental study has verified the scalability and efficiency of our methods. We conclude that partial evaluation provides a promising approach to distributed graph query evaluation.

We are currently developing distributed evaluation (MapReduce) algorithms for other queries, notably graph pattern matching, over larger real-life graphs. Another topic is to combine partial evaluation and incremental computation, to provide efficient distributed graph query evaluation strategies in the dynamic world.

**Acknowledgments**. Fan, Wang and Wu are supported in part by the 973 Program 2012CB316200 and NSFC 61133002 of China, and in part by EPSRC EP/J015377/1, an IBM scalable data analytics for a smarter planet innovation award, and the RSE-NSFC Joint Project Scheme.